\documentclass[preprint,12pt,longnamesfirst,epsf]{aastex}

\newcommand  \kms   {km~s$^{-1}$}

\newcommand  \ha    {H$\alpha$} 

\newcommand  \hii   {\ion{H}{2}} 
\newcommand  \hi    {\ion{H}{1}}
\newcommand  \nii   {[\ion{N}{2}]}

\shorttitle{Clusters in M101 Giant \hii\ Regions}
\shortauthors{Chen, Chu, \& Johnson}

\begin{document}

\title{Clusters in the Luminous Giant \hii\ Regions in M101}

\author{C.-H. Rosie Chen and You-Hua Chu}
\affil{Department of Astronomy, University of Illinois, 1002 
 West Green Street, Urbana, IL 61801}
\email{c-chen@astro.uiuc.edu, chu@astro.uiuc.edu}

\and

\author{Kelsey E. Johnson\altaffilmark{1,2}}
\affil{Department of Astronomy, University of Wisconsin, 475 North 
 Charter Street, Madison, WI 53706}
\email{kjohnson@astro.wisc.edu}

\altaffiltext{1}{Hubble Fellow and NSF Astronomy and Astrophysics 
 Postdoctoral Fellow.}
\altaffiltext{2}{present address: Department of Astronomy, University
 of Virginia, P.O. Box 3818, Charlottesville, VA 22903;
 kej7a@virginia.edu.}

\begin{abstract}

We have obtained {\it HST} WFPC2 observations of three very 
luminous but morphologically different giant \hii\ regions 
(GHRs) in M101, NGC\,5461, NGC\,5462 and NGC\,5471, in order
to study cluster formation in GHRs.
Images obtained in the F547M and F675W bands are used to 
identify cluster candidates and for photometric measurements,
and images in the F656N band are used to show ionized 
interstellar gas.
The measured colors and magnitudes are compared with the 
evolutionary tracks generated by the Starburst99 and Bruzual 
\& Charlot population synthesis models to determine the ages
and masses of the cluster candidates that are more luminous 
than $M_{\rm F547M}=-9.0$.
The brightest clusters detected in the PC images are measured 
and found to have effective radii of 0.7--2.9 pc.

NGC\,5461 is dominated by a very luminous core, and has been
suggested to host a super-star cluster (SSC).
Our observations show that it contains three R136-class clusters
superposed on a bright stellar background in a small region.
This tight group of clusters may dynamically evolve into an SSC 
in the future, and may appear unresolved and be identified as 
an SSC at large distances, but at present NGC\,5461 contains
no SSCs.
NGC\,5462 consists of loosely distributed \hii\ regions and
clusters without a prominent core.
It has the largest number of cluster candidates among the 
three GHRs studied, but most of them are faint and older 
than 10 Myr.
NGC\,5471 has multiple bright \hii\ regions, and contains
a large number of faint clusters younger than 5 Myr.
Two of the clusters in NGC\,5471 are older than R136, but 
just as luminous; they may be the most massive clusters
in the three GHRs studied.

The fraction of stars formed in massive clusters has been
estimated from the clusters' contribution to the total
stellar continuum emission and from a comparison between
the ionizing power of the clusters and the ionizing requirement
of the associated \hii\ regions.
Both estimates show that $\lesssim$ 50\% of massive 
stars are formed in massive clusters; consequently, the \ha\ 
luminosity of an \hii\ region does not provide a sufficient 
condition for the existence of SSCs.
The cluster luminosity functions (LFs) of the three GHRs show
different slopes.
NGC\,5462 has the steepest cluster LF and the most loosely 
distributed interstellar gas, qualitatively consistent with
the hypothesis that massive clusters are formed in high-pressure
interstellar environments.
The combined cluster LF of the three GHRs has a slope similar to 
the universal cluster LFs seen in starburst galaxies and 
non-starburst spiral galaxies.

\end{abstract}


\clearpage

\section{Introduction}

Giant \hii\ regions (GHRs) are sites of intense massive star formation. 
Their \ha\ luminosities, $10^{39}$--$10^{41}$ ergs~s$^{-1}$ \citep{Ke84},
require an ionizing power equivalent to that of 24--2400 O5V stars 
\citep{SdK97}.
With such high concentrations of massive stars, GHRs provide an
excellent laboratory to study the modes of massive star formation,
and in particular to probe whether they are birthplaces of globular 
clusters \citep{KC88}.

In the two nearest GHRs, 30 Dor in the Large Magellanic Cloud 
(LMC) and NGC\,604 in M33, where stellar contents are well resolved,
two distinct types of stellar groupings have been observed: 30 Dor 
is dominated by one central massive cluster R136 \citep{Hu95,WB97}, 
while NGC\,604 contains multiple OB associations spreading over
a large area \citep{Hu96}.
Evidently, not all GHRs contain massive compact clusters;
what physical environments give rise to the various cluster 
morphologies is currently under investigation.

One obvious way to elucidate this issue is to carry out detailed 
examinations of relatively nearby clusters and their environments.
\citet{MA01} studied 27 nearby ($<$ 5 Mpc) clusters of varying 
morphological types, primarily classifying clusters based on their
core and halo sizes.  
He suggests that compactness of clusters is predominantly related to 
the central density of the progenitor giant molecular cloud, i.e., 
extremely high pressure environments may be required to form
massive compact clusters.
However, this scenario has not been observationally tested; we do 
not know the pressures and densities of the giant molecular cloud 
in which optically visible clusters were formed.  
Examining clusters in a range of present-day environments may help 
us gain insight into their properties and relationship to their 
natal interstellar medium.

Of all massive compact clusters, the most impressive ones are the
super-star clusters (SSCs) with masses of 10$^5$--10$^6$ M$_\odot$.
SSCs are frequently observed in galaxy mergers and starburst galaxies 
\citep[][and references therein]{Wh03}, and they are believed to be 
preferentially formed in high-pressure interstellar conditions
\citep[e.g.,][]{EE97}.
However, some GHRs in non-interacting, late-type spiral galaxies 
may also host SSCs, especially those GHRs that are several times 
as luminous as 30 Dor and require ionizing powers rivaling those 
of young SSCs \citep[e.g.,][]{LP01}.
It is thus intriguing to examine the cluster content of such
GHRs and investigate whether these relatively quiescent environments 
can also produce SSCs.

The giant spiral galaxy M101 contains a large number of very
luminous GHRs whose stellar content can be resolved and studied
with {\it Hubble Space Telescope} ({\it HST}) images.
We have therefore obtained {\it HST} observations of three M101 
GHRs with different morphologies and galactic locations: NGC\,5461, 
NGC\,5462, and NGC\,5471.
The locations of these GHRs in M101 are marked on a Second Palomar 
Observatory Sky Survey (POSS-II) red image in Figure~\ref{fig:m101}.
The properties of these GHRs are summarized in Table~\ref{tbl:GHRs};
for comparison, 30 Dor is also included in this table.
We have used the {\it HST} continuum and \ha\ images of these three 
GHRs to carry out a detailed photometric study of their clusters.
This paper reports our observations (\S2) and methodology (\S3),
describes the cluster content in each GHR (\S4), discusses cluster
properties and their implications in studying massive star formation 
and cluster formation (\S5), and summarizes our results (\S6).

\section{Observations and Data Reduction}

The {\it HST} WFPC2 images of the GHRs NGC\,5461, NGC\,5462, 
and NGC\,5471 were obtained for the Cycle 6 program GO-6829.  
The observations were made through the continuum filters 
F547M (Str\"omgren $y$) and F675W (WFPC2 $R$), 
and the \ha\ filter F656N \citep[for filter
characteristics, see][]{Bi96}.  
Multiple exposures in each filter were made with a GHR centered 
on a Wide Field Camera (WFC) for photometric measurements.
Two short exposures in F547M with the GHR centered on the 
Planetary Camera (PC) were also made for cluster size measurements.
The observations are listed in Table~\ref{tbl:log}.
 
We received the {\it HST} pipeline processed WFPC2 images and 
then reduced them further with the IRAF and STSDAS routines.
All images were corrected for the intensity- and position-dependent 
charge transfer efficiency (CTE) by applying a linear ramp with a 
correction factor chosen according to the average counts of the sky 
background \citep{Ho95}.  
Images in each filter were then combined to remove cosmic rays and to 
produce a total-exposure map.  
To better illustrate the spatial correlation between the stars/clusters
and the ionized gas, we have produced color images of the three GHRs
using a customized IDL routine.
The individual F547M, F675W, and \ha\ images were mapped 
to the colors blue, green, and red, respectively.  
These images were transformed to a logarithmic scale, and the maximum 
and minimum flux values for each filter were chosen in order to 
maximize the dynamic range of the image while also creating a 
relatively black background.
The color images of NGC\,5461, NGC\,5462, and NGC\,5471 are shown in
Figures~\ref{color1}--\ref{color3}, where the ionized gas appears red 
and most stars blue.
The individual F547M, F675W, and \ha\ images of NGC\,5461, 
NGC\,5462 and NGC\,5471 are presented in 
Figures~\ref{fig:n5461}--\ref{fig:n5471}, respectively.

Aperture photometry was carried out using the IRAF task APPHOT
for the F547M and F675W images. 
Owing to the small number of identifiable candidate sources and
the complex blending and irregular background in some regions,
we manually selected compact sources with obvious peaks as 
candidate clusters in the three GHRs.
The candidate clusters in the three GHRs are marked in 
Figures~\ref{fig:n5461}--\ref{fig:n5471}.
The apparent magnitudes, $m_{\rm F547M}$ and $m_{\rm F675W}$, 
were measured with the WFC images using a source aperture of 
radius 2 pixels (0\farcs2) and an annular background aperture of 
radii 6--11 pixels.
For clusters with neighboring clusters within $\lesssim$ 0\farcs3,
such as \#8, \#9 and \#10 in NGC\,5461 and \#1 and \#2 in NGC\,5471,
the photometry was measured with a 0\farcs15-radius 
source aperture using both the WFC and PC images.
The corrections from the 0\farcs15-radius aperture to the 
0\farcs2-radius aperture are determined by measuring isolated 
resolved and unresolved sources in each image, and are in the
range of $-0.2$ to $-0.3$ mag.
The fluxes of these clusters will be over-estimated from the
WFC images due to the inclusion of neighbor's light, and the 
errors are larger for the fainter clusters.
For example, the error in the $m_{\rm F675W}$ of NGC\,5461-9 may 
be as large as $-$0.4 mag, which is estimated by comparing its
$m_{\rm F547M}$ measured from the PC and WFC images.

We have derived the magnitudes in the VEGAMAG system.  
The errors in $m_{\rm F547M}$ given by APPHOT are $\sim 0.01$ mag 
for $m_{\rm F547M} \le$ 20, and rise to 0.02--0.03 mag for 
$m_{\rm F547M} =$ 20--21.  
The errors in $m_{\rm F675W}$ are generally larger, with 
most of them $\le 0.02$ mag but some as high as 0.04 mag.  
These formal errors are derived from the flux variations 
in the background annulus used in the photometric measurements.
In regions with bright irregular backgrounds, using different
background apertures may produce different photometric results 
and the uncertainties in photometry will be larger than the
formal errors given by APPHOT.  
We have taken these uncertainties into account and estimated 
realistic errors.

The bright irregular sky background is attributed to both stars 
and nebulosity.  
The extended distribution of unresolved stars, similar to the 
``star clouds'' defined by \citet{LH70} for OB associations in 
the LMC, raises the background in both the F547M and 
F675W images.  
We find that using different annular sky backgrounds results in 
uncertainties of $\sim$ 0.03--0.05 mag in $m_{\rm F547M}$ for sources 
near modest star clouds, $\sim$ 0.1 mag for sources surrounded by 
bright star clouds, and up to $\sim$ 0.2 mag for faint sources near 
bright star clouds.  
The stellar background does not affect the ($m_{\rm F547M} - 
m_{\rm F675W}$) color as much because the variations in $m_{\rm F547M}$ 
and $m_{\rm F675W}$ are correlated.  
The bright nebular background, on the other hand, contributes 
uncertainties only to $m_{\rm F675W}$ and hence affects the color. 
The uncertainties in the ($m_{\rm F547M}-m_{\rm F675W}$) color are
$\sim$ 0.05--0.06 mag for most sources near nebulosities, and up to 
$\sim$ 0.16 mag for faint stellar sources near bright nebulosity. 

To reduce the uncertainties introduced by nebular contamination 
in the F675W images, we have produced \ha-free F675W images 
by subtracting scaled \ha\ images from the F675W images.
The \ha-subtracted F675W images of the three GHRs are also
presented in Figures~\ref{fig:n5461}--\ref{fig:n5471}.
Aperture photometry has been carried out for the \ha-subtracted 
F675W images, and the apparent magnitude is designated as 
$m_{\rm F675W'}$.  
The uncertainties in the ($m_{\rm F547M}-m_{\rm F675W'}$) color with 
different annular sky backgrounds are reduced to $\sim$ 0.02--0.05 mag 
for most candidate cluster sources near nebulosities.  

Our 0\farcs2-radius source aperture does not include all the light 
from a cluster.
The correction from a 0\farcs2-radius aperture to a 0\farcs5-radius 
aperture, which includes $\sim$ 95\% of the light of a point source
\citep{Ho95}, is $\sim -0.20\pm0.06$ mag for unresolved sources
and larger for resolved sources.
As the cluster candidates are resolved to different extents, the 
aperture corrections are in the range of $-$0.2 to $-$0.3 mag but
difficult to determine exactly.
We have chosen not to apply aperture corrections; therefore, our
photometric measurements are systematically fainter by 0.2--0.3 mag, 
but the analysis and conclusions of this paper are not sensitive to 
such small offsets that incur on the magnitudes and colors of 
the clusters.

The photometric results of NGC\,5461, NGC\,5462 and NGC\,5471 are 
presented in Tables~\ref{tbl:photos-1}--\ref{tbl:photos-3}, and plotted 
in the color-magnitude diagram (CMD) of $M_{\rm F547M}$ versus 
($M_{\rm F547M} - M_{\rm F675W'}$) in 
Figures~\ref{fig:cmd_5461}--\ref{fig:cmd_5471}, respectively.
For clusters with close neighbors, their $M_{\rm F547M}$ measured from
the PC images and their ($M_{\rm F547M} - M_{\rm F675W'}$) measured from
the WFC images are used in the CMD.
These absolute magnitudes are derived using a distance modulus 
of $(m-M) = 29.3$ \citep{St98}. 
The Galactic foreground extinction toward M101, $E(B-V)=0.01$
\citep{Sc98}, is corrected, although its effect is negligible.
The internal extinction from M101 is not individually corrected for, 
given that it is highly variable and the measurements are only
available for certain parts of the GHRs.

\section{Methodology}

The observed magnitudes and colors of the clusters can be used 
to determine their ages and masses through comparisons with 
those predicted by population synthesis models 
\citep[e.g.,][]{EF85,BC93,BC03,Le99}.
Below we describe the synthetic photometry derived from models
and how we use it to estimate the properties of clusters.
We have also used the \citet{La99} method to measure the sizes of 
clusters from their surface brightness profiles.
The procedures of cluster size measurements are outlined at 
the end of the section.

\subsection{Synthetic Photometry}

We have used the Starburst99 models \citep{Le99} and the 
\citet[][hereafter BC03]{BC03} models to generate synthetic 
photometry for comparison with observations of our cluster 
candidates in GHRs.
We have adopted a Salpeter initial mass function (IMF) with lower 
and upper mass limits of 1 M$_{\odot}$ and 100 M$_{\odot}$, which 
are commonly used in population synthesis models for star-forming 
and starburst regions. 
The luminosity, colors, and evolution of a cluster depend on
its metallicity.
To select appropriate models, we have used the observed oxygen 
abundances of the GHRs to assess their metallicities, because 
clusters and their surrounding GHRs are expected to have the 
same abundances and the oxygen abundances are well determined.
Oxygen abundances of NGC\,5461 and NGC\,5462, relative to the solar 
value, have been measured to be 0.6--0.9, while that of NGC\,5471 
is $\sim$ 0.25 \citep[e.g.,][]{Ev86,Sc92,Pi01,Lu02}.
Therefore, we adopt the 1 $Z_\odot$ model for NGC\,5461 and 
NGC\,5462, and the 0.2 $Z_\odot$ model for NGC\,5471.

The F547M and F675W filters we used are not included in the
default filter systems of Starburst99 or BC03 for which synthetic
photometry is readily available; thus, customized procedures are 
needed to derive synthetic $M_{\rm F547M}$ and $M_{\rm F675W'}$.  
As the first step, we use Starburst99 Version 4.0 to generate 
integrated stellar spectra for a simple stellar population (SSP; 
i.e., a single-age and single-abundance group of stars) from ages 
of 0 to 30 Myr at 1 Myr intervals and from 30 to 150 Myr at 3 Myr 
intervals.
This step is not necessary for the BC03 models, as integrated 
spectra for an SSP are available for most of these age intervals.
These model spectra are those without nebular line and continuum
emission because the clusters are generally well resolved from the
superposed extended nebular emission and the
background-subtraction in APPHOT adequately removes 
the extended nebular emission.
The synthetic spectra from Starburst99 and BC03 are then 
convolved with filter transmission curves, using the IRAF/STSDAS 
task CALCPHOT, to calculate the synthetic $M_{\rm F547M}$ and 
$M_{\rm F675W'}$.
We have produced synthetic photometry for SSPs with metallicities of 
0.2 $Z_\odot$ and 1 $Z_\odot$, and generated evolutionary tracks in the
CMDs in Figures~\ref{fig:cmd_5461}--\ref{fig:cmd_5471} for comparisons 
with observations of NGC\,5461, NGC\,5462, and NGC\,5471, respectively.
The differences in the two sets of evolutionary tracks reflect the
differences between the Geneva and Padova stellar evolution models
used by Starburst99 and BC03, respectively.
However, the effects of these differences are small compared to the 
uncertainties in the cluster mass estimates.

We use the R136 cluster at the core of 30 Dor as a reference
point, because it is an archetypical populous blue cluster and
possibly a young globular cluster.
The R136 cluster is $\sim$ 3 Myr old \citep{Hu95,WB97}.
Its spatially resolved photometry in the $V$ band has been measured
and the absolute visual magnitude within a radius of 7 pc is 
$M_V$ = $-$11.1 \citep{Mo85}.
This $V$ band magnitude is adopted directly because its central
wavelength is similar to that of F547M and the 7-pc-radius
aperture matches that used in the photometric measurements of
M101 clusters.
The ($M_{\rm F547M} - M_{\rm F675W'}$) color of R136 is not available,
so we use the synthetic color generated by Starburst99 for a 3 Myr
old cluster with $Z$ = 0.2--0.4 $Z_\odot$.
As no extinction correction has been applied to the M101 clusters,
we have reddened the synthetic color of R136 with its visual extinction
$A_V = $ 1.2 \citep{Mo85} and marked both the dereddened and reddened 
R136 in the CMDs in Figures~\ref{fig:cmd_5461}--\ref{fig:cmd_5471}.

\subsection{Assessing Masses and Ages of Clusters}

The mass and age of a cluster can be assessed by comparing 
its magnitudes and colors to model predictions if photometric 
data are available in three passbands.
For a young cluster, it is important to include a $U$ band or 
a $B$ band because the spectral energy distribution of young
massive stars peaks in the ultraviolet wavelengths.
Unfortunately, only F547M and {\rm F675W} photometry 
is available for the clusters in M101 GHRs, and these two bands 
are not as sensitive to the young massive stars as the $U$ and $B$
bands.
We cannot determine unambiguously the cluster masses and ages 
by comparing the photometric measurements with the evolutionary 
tracks of SSPs in the CMDs.
However, we may use the interstellar environment as an independent 
diagnostic of the cluster age, as the interstellar medium around 
a cluster evolves as a result of stellar energy feedback.

At ages $<$ 5 Myr, a cluster has the highest ionizing power, 
and hence will be in a dense, luminous \hii\ region.  
At ages 5--10 Myr, the fast stellar winds and supernova explosions
from a cluster have swept up the ambient ISM into a supershell 
with a visible cavity around the cluster.  
At ages $>$ 10 Myr, a cluster loses its ionizing power and has 
dispersed its ambient gas, so it will be surrounded only by diffuse
gas with low surface brightness.
We have compared the \ha\ images with the continuum images to 
examine the interstellar environment of the clusters and to assess 
the approximate ages of the clusters.
In Figures~\ref{fig:cmd_5461}--\ref{fig:cmd_5471}, we mark circles
around the clusters that are coincident with compact, luminous \hii\ 
regions, and dashed circles around the clusters that are in supershells,
indicating that their ages are $<$ 5 Myr and 5--10 Myr, respectively.
The unmarked clusters, not surrounded by bright \ha\ emission,
are older than 10 Myr, but their exact ages are poorly constrained.

With a rough estimate of the cluster age, it is then possible to 
compare the location of a cluster in the CMD with the synthetic
evolutionary tracks of clusters to determine the cluster mass.
The photometric measurements of the clusters have not been 
corrected for the extinction within M101, thus using these reddened 
magnitudes would underestimate cluster masses.
To illustrate the effect of extinction and to make a rough correction,
we take the visual extinctions of the GHRs determined from their 
Balmer decrements by \citet{KG96}, and plot the corresponding 
reddening vectors in Figures~\ref{fig:cmd_5461}--\ref{fig:cmd_5471}.
We have adopted these nebular extinctions and made reddening-corrected 
estimates of masses for clusters more luminous than $M_{\rm F547M} = -9$.
The age and mass estimates of these luminous clusters are 
given in Table~\ref{tbl:mass}.
We do not attempt to estimate masses for fainter clusters because 
these luminosities overlap those of single supergiants \citep{HD79}.
Furthermore, many faint clusters are not surrounded by bright
nebulosity, indicating poorly constrained ages at $>$10 Myr, 
so their mass estimates would be highly uncertain.

We have used the R136 cluster to estimate the uncertainties in 
our cluster mass estimates.
From the extinction-corrected location of R136 in the CMD, we
estimate a mass of $\sim 2 \times 10^4$ M$_\odot$.
The mass of R136 has been derived from its resolved stellar 
content to be 2.2 $\times 10^4$ M$_\odot$ \citep{Hu95} by
summing the masses of stars $\ge$ 2.8 M$_\odot$ (mass cutoff
limited by completeness) within a 4.7-pc radius.
Note however that our estimate of mass is based on the luminosity of 
R136 within a 7-pc radius and a minimum stellar mass of 1 M$_\odot$.
Our mass estimate of R136 using the 4.7-pc radius aperture \citep{Mo85} 
and the 2.8 M$_\odot$ lower mass limit is 1.4 $\times 10^4$ M$_\odot$, 
about 40\% lower than that derived from the resolved stellar content.
Therefore, the uncertainties in our cluster mass estimates are
at least 40\%.

\subsection{Assessing Cluster Sizes}

Some of our clusters appear resolved in the PC images, so it is 
possible to determine their sizes.
The size of a cluster can be described by its effective radius, 
$R_{\rm eff}$, the radius that encircles half of the cluster light.
The $R_{\rm eff}$ of a cluster can be estimated with the routine 
ISHAPE developed by \citet{La99}.
In this routine, the surface brightness profile of a cluster is 
modeled by an analytic function and convolved with a point spread 
function (PSF) calculated with the TINY TIM Version~6.0 \citep{Kr95} 
for the cluster's position on the PC chip.
The PSF-convolved model profile is then compared with the observed
cluster profile.
The best-fit model, judged by the $\chi^2$ statistics, gives 
the full width at half maximum (FWHM) of the analytic function,
which is then used to determine $R_{\rm eff}$.

We have used the two most common analytic functions, the
Gaussian and the \citet{Ki62} profiles, to model the clusters.
The King profile contains a concentration parameter
$c =$ log($r_{\rm t}/r_{\rm c}$), where $r_{\rm t}$ is the tidal 
radius and $r_{\rm c}$ is the core radius.
Typically $c$ is within the range of 1.0--2.0 for globular clusters 
in the Galaxy \citep{Ha96} and young rich clusters in the LMC 
\citep{EFF87}.
We have experimented with different values of $c$ within this 
range in the model fits, and found that for a cluster detected 
with $S/N \ge$ 15, the best-fit $R_{\rm eff}$ is insensitive to the 
concentration parameter $c$ or even the form of the analytic function.
For a bright cluster with adequate $S/N$, the $R_{\rm eff}$ is 
estimated using both the Gaussian and the King profiles, and the
average of the two estimates is adopted and given in 
Table~\ref{tbl:mass}.

All but one of the cluster sizes we measured are in the range 
of $R_{\rm eff}$ = 0\farcs02--0\farcs09, corresponding to
0.7--2.9 pc. 
(Note that $R_{\rm eff}$ can be smaller than the pixel
size of 0\farcs0455, because the PSF effects have been
considered and removed in the profile fitting.)
The only exception is NGC\,5471--9, whose $R_{\rm eff}$ is only 
0\farcs005; as we discuss later in \S4.3, NGC\,5471--9 is most 
likely a luminous A-F supergiant, instead of a cluster.
The sizes of these M101 clusters are within the range of clusters 
in the Galaxy and nearby galaxies.
For example, the globular clusters in the Galaxy have 
$R_{\rm eff} \sim$ 1--5 pc, with a median of $\sim$3 pc \citep{Ha96}; 
the compact young cluster R136 in the LMC has 
$R_{\rm eff} \sim$ 0.9 pc \citep{MG03}\footnote{The surface brightness 
profile of R136 indicates a compact, dominant component on top of a 
broad, shallow component. The $R_{\rm eff}$ is estimated using its 
core radius of 0.32 pc and a King profile with $c = 1.5$ to approximate 
the compact component.}; compact young massive clusters in nearby 
starburst galaxies where stars are not resolved have $R_{\rm eff} \sim$ 
2--4 pc \citep{Metal95}.
Comparisons between the M101 clusters and R136 will be discussed 
in more detail in \S4.

\section{Clusters in Three Luminous GHRs in M101}

Below we describe the spatial distribution, ages, masses, and sizes
of the clusters in NGC\,5461, NGC\,5462, and NGC\,5471.
The extinctions of individual \hii\ regions in these GHRs are taken 
from \citet{KG96}.

\subsection{Clusters in NGC\,5461}

The GHR NGC\,5461 has been loosely defined to be the \hii\ complex 
extending over a $66''\times26''$ region in ground-based \ha\ 
images \citep{Is75}.
Considering that this area corresponds to a linear size of
$\sim 2.3$ kpc $\times$ 0.9 kpc, it is unlikely that the entire
region is associated with one coherent star formation event.
Indeed, 12 \hii\ regions have been identified within NGC\,5461 
by \citet{Ho90}.
Our WFPC2 \ha\ image shows that NGC\,5461 contains two regions 
that would have been individually identified as GHRs if they
were in the Local Group: H\,1105 and H\,1098 
\citep[marked in Figure~\ref{fig:n5461}e; designation from][]{Ho90}.
H\,1105 is 3 times as luminous as 30 Dor, and H\,1098 is as 
luminous as NGC\,604, or 1/3 as luminous as 30 Dor.
The 10 fainter \hii\ regions are distributed roughly along
the axis connecting H\,1105 and H\,1098 with a higher concentration
toward H\,1105. 
We define the ``main body" of NGC\,5461 to be the region containing
H\,1105 and H\,1098 and their vicinity, as in the field-of-view 
of Figure~\ref{fig:n5461}.

The relationship between the stars/clusters and the \hii\ 
regions is clearly illustrated in the color image of NGC\,5461 
(Figure~\ref{color1}).
In addition to the bright \hii\ regions, NGC\,5461 also has 
nebular filaments, loops, and well-defined shells with 
stars/clusters underneath these interstellar structures. 
A total of 12 cluster candidates are identified in the main 
body of NGC\,5461; they are listed in Table~\ref{tbl:photos-1} 
and marked in Figure~\ref{fig:n5461}f.
Only six clusters have $M_{\rm F547M} \le -9$: five in H\,1105 
(\#6, \#8, \#9, \#10, and \#11) and one in H\,1098 (\#1).
A careful inspection of their immediate surroundings shows
that all six clusters are superposed on bright \hii\ regions, 
indicating ages of $<$ 5 Myr.
To estimate the masses of these young luminous clusters, we adopt
the visual extinctions of $A_V = 1.7\pm0.4$ and $0.8\pm0.1$ of 
H\,1098 and H\,1105, apply the respective extinction correction, 
and compare the dereddened cluster positions in the CMD in 
Figure~\ref{fig:cmd_5461} with the evolutionary tracks generated
by Starburst99 and BC03 for different SSP masses.
Four of the clusters, \#1, \#6, \#9, and \#10, show dereddened
colors consistent with SSPs at ages $<$ 5 Myr; thus their masses 
can be estimated in a straightforward manner.
The two remaining clusters, on the other hand, have dereddened
colors consistent with SSPs at ages greater than 6 Myr.
The red colors of clusters \#8 and \#11 can be caused by large local 
extinction excesses or stochastic color deviation for low-mass SSPs.
We consider the latter more likely, i.e., the cluster contains or 
is projected near a red supergiant and the cluster color is thus 
confused.
For example, cluster \#11 may consist of (or be projected towards) a
K0 Ia-O supergiant with $M_V=-9.4$ \citep{Hu78} and (V-R)$_0=0.76$ 
\citep{Jo66} and a young cluster with $M_{\rm F547M} = -9.2$.
The masses of clusters \#8 and \#11 are determined with the ad hoc
assumption of a contaminating red supergiant, which reduces
the luminosity of the cluster and lowers the mass estimate 
accordingly.
The mass estimates of the six brightest clusters in NGC\,5461
are mostly in the range of 1--3 $\times10^4$ M$_\odot$
(see Table~\ref{tbl:mass}).
These masses are comparable to that of R136, $\sim 2 \times10^4$ 
M$_\odot$.

The PC images of NGC\,5461 are used to determine the cluster
sizes, but only three clusters, \#6, \#8, and \#10, are detected 
with $S/N \ge $15 for reliable size measurements.
The $R_{\rm eff}$ estimated for these three clusters are 0.8, 0.7, 
and 2.1 pc, respectively.
While the sizes of clusters \#6 and \#8 are comparable to that 
of R136, $R_{\rm eff} = 0.9$ pc, cluster \#10 is more extended 
and shows visible departure from spherical symmetry in the
PC image in Figure~\ref{fig:5461core}.
The morphology of cluster \#10 suggests that it may be a 
composite of two clusters with the southwest object brighter
than the northeast object.

Among the three GHRs we studied in M101, NGC\,5461 is particularly
interesting because it is one of the most luminous GHRs in galaxies 
within 10 Mpc \citep{Ke84}.
Furthermore, the core of NGC\,5461 (i.e., H\,1105) has a remarkably 
high surface brightness with a peak emission measure of $4.4\times10^5$
cm$^{-6}$~pc, comparable to those of the most active starburst regions.
The \ha\ luminosity of H\,1105 implies an ionizing flux rivaling 
those of SSCs \citep{KC88,LP01}; thus, NGC\,5461 has been considered 
the most promising site in M101 where SSCs might be found.
However, our analysis shows that H\,1105 contains five R136-class
clusters, which are by no means in the same league as the SSCs with 
masses $\sim 10^5$--10$^6$ M$_\odot$ commonly found in starburst 
galaxies or mergers \citep[e.g.,][]{OC94,Wh99}.
The core of NGC\,5461 is nevertheless striking in its high 
concentration of stars in a small volume -- three clusters with 
a total mass of $6\times10^4$ M$_\odot$ in a region of $\sim$ 32 pc
(0\farcs9) across.
It is possible that these clusters are subclusters that will 
dynamically interact and merge into a cluster that has a mass more 
typical for SSCs.
As shown in the numerical simulations of \citet{BBV03}, the 
hierarchical fragmentation of giant molecular clouds naturally leads 
to the formation of subclusters that can merge to form the final 
stellar cluster.  
If H\,1105 were projected to a distance similar to that of the Antennae 
galaxies \citep[$\sim 20$~Mpc,][]{Wh99}, it would imitate a single 
cluster as shown in Figure~\ref{fig:5461core}b, and the combined
light of the clusters and the bright stellar background in a 
0\farcs2-radius aperture would have $M_{\rm F547M} = -13.0$, 
corresponding to a mass of $\sim 10^5$ M$_\odot$.
These results suggest that some young SSCs previously identified at 
distances of $\gtrsim 20$~Mpc may be tight groups of R136-class 
clusters as seen in the core of NGC\,5461.

\subsection{Clusters in NGC\,5462}

The GHR NGC\,5462 corresponds to a large \hii\ complex with a
dimension of $90''\times34''$, or 3.2 kpc $\times$ 1.2 kpc,
in ground-based \ha\ images \citep{Is75}.
Thirty-three \hii\ regions have been identified within NGC\,5462
\citep{Ho90}, but none are comparable to 30 Dor.
The overall morphology of NGC\,5462 consists of a few bright
\hii\ regions distributed along an axis from northeast to
southwest and fainter filaments and loops extending outwards
from this axis.
The two brightest \hii\ regions, H\,1170 and H\,1176, are each only 
comparable to NGC\,604, or 1/3 as luminous as 30 Dor; the others 
are much fainter.
The distribution of star formation in NGC\,5462 is apparently
not as concentrated as in NGC\,5461.
We define the ``main body'' of NGC\,5462 to be the region containing
H\,1176, H\,1170, H\,1159 and their vicinity, as in the 
field-of-view of Figure~\ref{fig:n5462}.

The color image of NGC\,5462 (Figure~\ref{color2}) shows a distinct 
offset between the ionized gas and concentrations of stars, 
suggesting that the star formation has proceeded from the southeast
to northwest.
A total of 25 loosely distributed cluster candidates are 
identified in NGC\,5462 (see Table~\ref{tbl:photos-2} and 
Figure~\ref{fig:n5462}f).
Most of these clusters are faint; only three clusters, \#6, \#18, 
and \#23, have $M_{\rm F547M} \leq -9$ and are analyzed for their 
masses.
While cluster \#18 is superposed on a bright \hii\ region H\,1176, 
indicating an age of $<$ 5 Myr, clusters \#6 and \#23 are not 
associated with any \hii\ regions or supershells, indicating 
ages $>$ 10 Myr.
For cluster \#18 the visual extinction $A_V = 0.9\pm0.4$ 
of the surrounding \hii\ region H\,1176 is adopted, and for the 
other two clusters the visual extinctions of their nearest \hii\ 
regions are adopted, i.e., $A_V = 0.6\pm0.2$ of H\,1159 for cluster 
\#6 and $A_V = 0.9\pm0.4$ of H\,1176 for cluster \#23.
We apply the respective extinction correction to each of the three 
bright clusters and compare their dereddened cluster positions in 
the CMD in Figure~\ref{fig:cmd_5462} with the evolutionary tracks.
The young cluster \#18 shows a dereddened color consistent with 
SSPs at ages of $<$ 5 Myr and the estimated mass is $\lesssim 1 
\times 10^4$ M$_\odot$.
The two older clusters \#6 and \#23 show dereddened colors consistent 
with SSPs at ages of $>$ 10 Myr; however, as their ages are poorly 
constrained, we only obtain their lower mass limits by assuming 
cluster ages of $\sim$ 10 Myr.
We note that unless the two clusters have ages $>$ 30 Myr, their
masses would be within a factor of 1.5 of the lower mass limits. 
The mass estimates of the three brightest clusters in NGC\,5462 are 
in the range of 1--2 $\times 10^4$ M$_\odot$ 
(see Table~\ref{tbl:mass}), comparable to the mass of R136.

In the PC images of NGC\,5462, only cluster \#6 is detected with 
$S/N \geq 15$ for size measurements.
The $R_{\rm eff}$ estimated for this cluster is 
2.3 pc, more extended than that of R136.
As shown in Figure~\ref{fig:clusters}, cluster \#6 has an asymmetric 
morphology elongated along the northwest and southeast direction, 
indicating a complex structure.

NGC\,5462 has the lowest \ha\ surface brightness among the three 
GHRs we studied in M101.
It has a larger number of clusters than the other two GHRs, but 
only three are R136-class clusters.
Furthermore, the clusters do not show obvious spatial concentrations,
in sharp contrast to those seen in NGC\,5461.
The combination of the low \ha\ surface brightness and sparse 
distribution of small-mass clusters suggests that the star formation 
and cluster formation is more spread-out and modest in NGC\,5462.

\subsection{Clusters in NGC\,5471}

The GHR NGC\,5471 extends over a diameter of $\sim 17''$, or
$\sim$ 600 pc.
Ground-based images of NGC\,5471 show five bright knots, which 
are designated as A, B, C, D, and E components by \citet{Sk85}
and have been called NGC\,5471A--E, respectively.
Our color image of NGC\,5471 (Figure~\ref{color3}) shows that
the A-, B-, C-, and E-components display bright \hii\ regions 
centered on clusters.
However, the D-component displays an offset between the \hii\ 
region and the clusters, which are located to the north and the 
east sides of a dark cloud, respectively; it is uncertain whether 
the clusters and the \hii\ region are physically associated.
The A-component is as luminous as 30 Dor, and the B-, C-, and 
E-components are comparable to or fainter than NGC\,604.

A total of 19 cluster candidates are identified in NGC\,5471; they
are listed in Table~\ref{tbl:photos-3} and marked in 
Figure~\ref{fig:n5471}f.
Most of the clusters reside within the A--E components, with 
the highest concentration located in the A-component and its 
western extension.
The eight clusters with $M_{\rm F547M} \le -9$ have been analyzed. 
We have adopted the visual extinction of each component, applied
the respective extinction correction to the eight brightest clusters,
and compared the dereddened cluster positions in the CMD with the 
evolutionary tracks (see Figure~\ref{fig:cmd_5471}).
The comparisons are problematic because most of the dereddened cluster 
colors do not agree with those expected from the evolutionary tracks.
While the disagreements are partially attributed to errors in the 
extinction and photometric measurements, the dominant cause of the
disagreements is probably uncertainties in the stellar evolution 
models at low metallicities.
It is known that at $Z \leq 0.2~Z_\odot$ stellar evolution models 
cannot reproduce the observed luminosities and colors of red 
supergiants or the number ratios of blue to red supergiants
\citep[][]{Ma97,Oetal99,Le99}.
These uncertainties directly affect the luminosities and colors
of SSPs, particularly at ages around 7 to 14 Myr when red supergiants 
are significant contributors of the total light.
Given these uncertainties, we can make only order-of-magnitude 
mass estimates for the clusters in NGC\,5471.

Four of the clusters we analyzed (\#2, \#7, \#12, and \#16) are 
superposed on bright \hii\ regions, suggesting that they are 
$<$ 5 Myr old.
Clusters \#7, \#12, and \#16 are only $\sim$ 0.1 mag bluer or redder
than those expected for young clusters; therefore, we disregard these
color differences and use only the $M_{\rm F547M}$ to estimate their
cluster masses.
Cluster \#2, on the other hand, is $\sim$ 0.6 mag redder than the
color expected for its young age, and this discrepancy is larger
than the known errors in photometry or stellar evolution models.
We suggest that this red color excess is likely attributed to a 
contaminating post-outburst luminous blue variable (LBV) because 
cluster \#2 is located in the C-component where high-velocity 
($>$ 1000 km~s$^{-1}$), \nii-bright nebular emission similar to 
that of $\eta$ Car's ejecta nebula has been reported \citep{CVC90}.
We assume that cluster \#2 contains an LBV similar to $\eta$ Car, 
which has $V=6.22$ and $R=4.90$ at quiescent states \citep{Me67} and 
can brighten up by 1--2 mag during outbursts or 3--5 mag during 
super-outbursts \citep{HD94}.
These quiescent $V$ and $R$ magnitudes can be converted to 
$M_{\rm F547M} = -8.5$ and $M_{\rm F675W} = -9.4$ using the distance 
modulus of $(m-M)_0=12.79$ and the extinction of $A_V = 1.92$ for 
$\eta$ Car's host cluster Trumpler 16 \citep{DE01}.
The remaining members of cluster \#2 would have 
($M_{\rm F547M} - M_{\rm F675W}$) = 0.44, which is still too red
for a $< 5$ Myr old cluster.
However, if the hypothesized LBV is 0.4 mag brighter (during or 
after an outburst), the rest of cluster \#2 would have 
($M_{\rm F547M} - M_{\rm F675W}$) = $-0.1$ as expected for a young 
cluster and $M_{\rm F547M}$ = 8.2. 
These resultant color and magnitude are used to estimate the mass 
of cluster \#2.

The other four clusters we analyzed (\#3, \#4, \#5, and \#9) are not 
associated with \hii\ regions or supershells, suggesting that their 
ages are $>$ 10 Myr.
However, as discussed in the next paragraph, \#9 has a small intrinsic
size which makes it more likely a luminous supergiant rather than 
a cluster.
Among the remaining clusters, \#4 shows a dereddened color consistent 
with SSPs at ages $>$ 10 Myr and thus its lower mass limit is easily 
estimated.
Clusters \#3 and \#5, on the other hand, have dereddened colors
$\sim 0.2$ mag bluer than that of SSPs at ages $>$ 10 Myr.
Since their extinction corrections are already small, the disagreements
are unlikely to be caused by the uncertainty in extinction measurements.
It is most likely that the disagreements arise from the uncertainties
in the modeled colors and photometry, so we disregard the blue color 
excesses when estimating the cluster masses.
The mass estimates for the eight bright clusters are in the range of
$\sim$ 0.5--2 $\times 10^4$ M$_\odot$ (see Table~\ref{tbl:mass}), 
approaching or comparable to R136.

The cluster sizes are determined using the PC images of NGC\,5471.
Four clusters, \#4, \#5, \#9, and \#16, are detected with $S/N \ge $15.
The $R_{\rm eff}$ estimated for these four clusters are 
2.9, 1.4, 0.2, and 1.1 pc, respectively.
The small size of cluster \#9 suggests that it is either an unresolved
star or a post-core-collapse globular cluster \citep{Ha96}.
Post-core-collapse globular clusters are at least $10^9$ yr old
\citep[e.g.,][]{He85} and thus unlikely to exist in GHRs.
We consider it more likely that cluster \#9 is a star with 
magnitudes and colors compatible to those of a luminous A-F 
supergiant \citep{MF77,HMF90}.
For the three resolved clusters, \#16 has a size comparable to that 
of R136, while \#4 and \#5 are more extended and show asymmetric,
elongated morphologies in the PC image in Figure~\ref{fig:clusters}.

NGC\,5471 has a large number of young clusters with ages $< 5$ Myr.
However, the majority of these young clusters are faint with 
$M_{\rm F547M} \ge -9$, which may be small clusters with masses 
of a few $\times 10^3$ M$_\odot$ or just luminous supergiants.
Among the cluster candidates in NGC\,5471, \#4 and \#5 are the most 
massive ones since they are older than 10 Myr and still as luminous
as the young R136 cluster, suggesting that their masses are
higher that that of R136.

\section{Discussion}

\subsection{Nature of Faint Cluster Candidates in GHRs}

A large number of cluster candidates have been identified in 
the three M101 GHRs, but the nature of the faintest objects 
is uncertain.
It is possible that some of these faint cluster candidates
consist of multiple OB associations as observed in the nearby 
GHR NGC\,604 \citep{Hu96} and some are simply luminous
supergiants frequently seen near high concentrations of massive
stars, such as 30 Dor \citep{WB97}.
We have therefore simulated WFPC2 images of 30 Dor and NGC\,604 
at a distance of 7.2 Mpc and searched for ``clusters'' using the
same criteria as we did for cluster candidates in the M101 GHRs.
The spurious clusters in 30 Dor and NGC\,604 can be identified 
because their resolved stellar contents are known.
The real and spurious clusters in these two GHRs can then be
compared with the cluster candidates in M101 to better assess
the nature of the latter.

To simulate a WFPC2 image of 30 Dor at 7.2 Mpc, we have used a 
green continuum ($\lambda_c$ = 5130 \AA, $\Delta \lambda$ = 155 \AA)
image from the Magellanic Clouds Emission-Line Survey 
\citep[MCELS,][]{Smith99}, and binned the data to 3.5 pc per pixel.
The resultant image is displayed in Figure~\ref{fig:ghrs}a.
Using the same identification criteria for clusters in M101 GHRs, 
the two brightest objects in 30 Dor will be selected as clusters:
R136 and R131.
While R136 is a bona fide cluster, R131 ($=$ HD\,269902) is an A0 
supergiant with $V = 10.0$ (corresponding to $M_V + A_V = -8.5$)
and hence a spurious cluster.
The mis-identification of R131 as a cluster bolsters our choice
of a luminosity cutoff of $M_{\rm F547M} = -9.0$ for cluster 
mass estimates (\S3.2), as the fainter cluster candidates may be 
single supergiants.
It is interesting to note that two other concentrations of stars 
in 30 Dor are not identified as clusters: the Hodge 301 cluster
\citep{H88} and the OB association LH\,99 \citep{LH70}.
The Hodge 301 cluster, with an age of $\sim$ 20--25 Myr and a 
mass of a few $10^3$ M$_{\odot}$ \citep{GC00}, is too faint to 
meet our cluster identification criteria.
LH\,99, on the other hand, is too distributed to mimic a cluster. 

A WFPC2 image of NGC\,604 at 7.2 Mpc is simulated with its archival 
WFPC2 F547M image.
Adopting a distance of 0.84 Mpc to M33 \citep{FWM91}, the data are
binned to 3.5 pc per pixel, and the resultant image is displayed 
in Figure~\ref{fig:ghrs}b.
The OB associations in NGC\,604 appear as small concentrations on 
top of an irregular stellar background.
The four brightest concentrations have $M_{\rm F547M}=-8.0$ to $-9.0$,
luminous enough to meet our identification criteria for M101 clusters.
However, most of the concentrations have irregular shapes, and the 
brightest one does not even have an obvious boundary. 
Therefore, the NGC\,604-type OB associations may mimic faint clusters 
at best, with luminosities rivaled by those of supergiant stars.

For a direct comparison between M101 GHRs and the simulated
30 Dor and NGC\,604 at 7.2 Mpc, Figure~\ref{fig:ghrs} displays 
their images in the same spatial and intensity 
scales\footnote{The green image of 30 Dor was taken in a similar
but not identical wavelength band.  The intensity scale of the 
30 Dor image is selected to match that of the F547M images as much 
as possible so that objects with similar magnitudes appear similar
in both F547M and green band images.} over a 350 pc $\times$ 350 
pc field-of-view.
It is immediately clear that the brightest cluster candidates 
in M101 GHRs are more luminous than R136 and are most likely 
bona fide clusters.
The nature of the cluster candidates fainter than $M_{\rm F547M} 
= -9.0$, marked in Figure~\ref{fig:ghrs}, is less obvious.
Some faint cluster candidates may be blue supergiants because they 
have sharp images and appear isolated, and the blue color excludes
the possibility of post-core-collapse clusters; examples of these
include NGC\,5461-7, NGC\,5462-20, and perhaps NGC\,5471-13 and 14.
Some faint cluster candidates may be OB associations because they
appear extended without a sharp boundary; examples of these include
NGC\,5461-5, NGC\,5462-13 and 15, and NGC\,5471-10, 11, and 17.

\subsection{The Fraction of Massive Stars Formed in Clusters}

It has been suggested that massive stars form preferentially 
in associations and clusters \citep[][and references therein]{SPH00}.
Our {\it HST} WFPC2 images of the M101 GHRs show that several 
young R136-class clusters are superposed on discrete regions
of unresolved stellar emission, e.g., clusters \#8, \#9, and \#10 
at the core of NGC\,5461 and cluster \#16 at the core of NGC\,5471A.
The unresolved stellar backgrounds are most likely star clouds
that contain field stars and loosely assembled associations.
The similarities in locations and colors suggest that the clusters
and the background star clouds are formed from the same episode of
star formation.
It is then interesting to determine the fraction of massive stars 
that are formed in R136-class clusters in these regions.
We have used two different methods to determine this fraction: one 
is based on the contribution of cluster light to the total light, 
and the other is based on a comparison between the ionization flux
expected from the clusters and the ionizing flux required by
the surrounding \hii\ region.

We have selected four regions for this analysis: two in NGC\,5461, 
one in NGC\,5462, and one in NGC\,5471.
These regions are listed in Table~7, and their close-up F547M
and \ha\ images are presented in Figure~\ref{fig:4regions}.
The F547M images show that the clusters in these four regions 
are all superposed on discrete bright diffuse stellar backgrounds,
and the \ha\ images show that all are at the cores of bright
compact \hii\ regions.
We have measured the total light from these four regions in both 
F547M and F675W$'$ bands using the apertures marked on the
F547M images in Figure~\ref{fig:4regions} and described in
column 3 of Table~7.
The background, determined from the median of an annular region 
outside the \hii\ region, has been subtracted, although it 
contributes to only 1--2\% of the total light.
For the clusters, we have applied aperture corrections of 
$\sim -0.2$ mag to our photometric measurements made with 
a 0\farcs2-radius APPHOT aperture to account for the missing light.
The cluster-to-total light ratios,  $L_{\rm cluster}/L_{\rm total}$
given in columns 4 and 5 of Table~7, are in the range of 0.25--0.5;
the uncertainties are dominated by the photometry and aperture
corrections of the clusters, as they are superposed on bright
local stellar background.
The light ratios are slightly larger in the F547M band than in 
the F675W band, because the clusters are 0.1--0.2 mag bluer than 
their diffuse stellar background.
This color difference can be caused by an age difference of 
a few Myr, assuming that the clusters and the underlying star 
clouds have the same initial mass function.
As the precise ages are unknown, we cannot model the star clouds
to determine their masses; therefore, the cluster-to-total light
ratio can be considered only as an approximation of the fraction
of massive stars formed in clusters.
In the four regions we analyzed, about 25--50\% of the massive
stars are formed in R136-class clusters.

The \ha\ images in Figure~\ref{fig:4regions} show that the \hii\
regions around the clusters have rather well-defined boundaries
where the surface brightness drops off sharply.
Such morphology suggests that the \hii\ regions are likely 
ionization-bounded, or optically thick to ionizing radiation.
We have measured the \ha\ fluxes\footnote{Owing to the $\sim300$ 
\kms\ redshift of M101, two corrections need to be considered.
First, the filter transmission of the red-shifted \ha\ line is 
$\sim93$\% of the peak transmission, thus the extracted \ha\ 
flux should be multiplied by a correction factor of 1.07.  
Second, the \nii$\lambda$6548 line is red-shifted into the \ha\ 
bandpass at $\sim91$\% of the peak transmission and needs to
be removed.  The \nii\ contamination, estimated from the
\nii/\ha\ ratios reported by \citet{KG96}, amounts to 1--3\% of 
the \ha\ flux in most cases.} of these four regions using the 
apertures marked on the \ha\ images in Figure~\ref{fig:4regions}
and described in column 6 of Table~7.
The continuum-subtracted \ha\ images are used for the flux
measurements, and extinction corrections are made.
Assuming a 10$^4$ K optically thick \hii\ region, the derived 
\ha\ luminosity, $L_{{\rm H}\alpha}$, can be used to determine 
the required ionizing luminosity, $Q({\rm H}^0)$, through the 
relation
\begin{equation}
Q({\rm H}^0) = 7.4\times10^{11}~L_{{\rm H}\alpha}~{\rm photons~s}^{-1},
\end{equation}
where $L_{{\rm H}\alpha}$ is in units of ergs~s$^{-1}$.
The resultant ionizing luminosities ($Q_{\rm HII}$) of the four 
regions are given in column 7 of Table~7.

The ionizing luminosity expected from the clusters at different ages 
can be calculated using the Starburst99 models.
The ages of the clusters in the four selected regions are $<$ 5 Myr,
as indicated by the associated bright \hii\ regions.
Unfortunately, during the first 5 Myr a cluster's ionizing luminosity
decreases rapidly, dropping from the maximum at $\sim$ 1 Myr to 
a factor of 5--7 lower at 5 Myr \citep{Le99}; therefore, the
uncertainty in cluster age directly propagates into the 
uncertainty in ionizing luminosity of a cluster.
We have adopted a cluster age of 3 Myr and calculated the expected 
ionizing luminosities of the clusters ($Q_{\rm cluster}$) and their 
ratios to those required by the surrounding \hii\ regions.
These results are given in columns 8 and 9 of Table 7.
These ratios, 0.2--0.6, can be viewed as a very crude approximation 
of the fractions of massive stars formed in clusters.
An interesting corollary of this result is that the ionizing
luminosity, or \ha\ luminosity, of an \hii\ region is not
a sufficient diagnostic for the existence of SSCs because the 
majority of stars may reside outside clusters.

In the four regions of star formation we considered, the fraction
of massive stars in clusters estimated from the cluster-to-total 
light ratio is, within the uncertainties, consistent with that
estimated from the cluster-to-\hii-region ionizing luminosity 
ratio -- no more than about half of the massive stars are formed 
in the R136-class clusters.
On the other hand, some R136-class clusters, such as \#4 and \#5 
in NGC\,5471D, are not superposed on a bright stellar background,
and constitute the dominant components in their associated episode 
of star formation.
The fraction of stars formed in clusters must cover a range, which 
varies according to the physical conditions of star formation.

\subsection{Interstellar Environments of the GHRs}

We examine the distribution of interstellar molecular clouds and 
\ion{H}{1} gas in the three M101 GHRs in order to gain insight on 
the cluster formation process.
Molecular CO observations of these GHRs have been made by 
\citet{GC99} using both single-dish telescopes and interferometers.
NGC\,5461 has the strongest CO emission among the three GHRs, 
and its CO peaks appear concentrated toward the peaks of the 
\ha\ emission \citep[see Figure~4 in][]{GC99}, where the massive
clusters are located.
The CO emission toward NGC\,5462 is detected in single-dish 
observations but not in interferometric observations, indicating
that the molecular gas is distributed over a scale larger than the
synthesized beam, $\sim$3$''$.
No CO emission is detected in NGC\,5471, which may be attributed
to its low metallicity.
In the two GHRs where CO is detected, the distribution of molecular
clouds is similar to that of clusters: concentrated in NGC\,5461
and distributed in NGC\,5462.

The distribution of \hi\ gas in M101 and its relation with
ionized gas have been reported by \citet[][see their 
Figures~1--2]{Setal00}.
Their \hi\ map shows that NGC\,5461 and NGC\,5462 are in the
same spiral arm.
Assuming a trailing arm, the offsets of \hi\ ridge downstream 
from the stars in NGC\,5461 and NGC\,5462 are consistent with
the expectations of star formation triggered by density waves
\citep{Ro69}.
NGC\,5471 has a concentration of \hi\ gas but the large-scale 
distribution of \hi\ show a complex inter-arm structure, which
may have resulted from tidal interactions during the last 10$^9$ yr
\citep[][and references therein]{Wetal97}.
As \hi\ gas can be produced by photodissociation of the natal
molecular clouds \citep{AAT85,AAT86,Setal00}, converted to
\hii\ by photoionization, and dispersed by fast stellar winds
and supernova explosions, the distribution of \hi\ does not
provide adequately pertinent information about the cluster formation.

To study the physical conditions for cluster formation, it is 
necessary to examine the interstellar environment of the youngest
clusters before the interstellar conditions have been altered
by stellar energy feedback.
The embedded young clusters that are observable in the infrared but
not yet in the optical wavelengths \citep[e.g.,][]{KJ99,TBH00,JK03}
provide promising locations to study the physical conditions of 
cluster formation.

\subsection{Cluster Luminosity Function}

The luminosity functions (LFs) of young compact clusters have been 
studied in various types of galaxies with different star formation 
rates as a means to gain insight into the cluster formation process.  
To first order, the measured LFs for young compact cluster systems 
in merging or starburst galaxies are remarkably universal, and can 
be approximated by a power-law of the form $dN(L)/dL \propto 
L^{\alpha}$, with the exponent $\alpha \approx -2 \pm 0.2$ 
\citep[see][and references therein]{Wh03}.
The cluster LFs for a sample of nearby non-starburst spiral 
galaxies also show similar $\alpha$, $-$2.0 to $-$2.4 \citep{La02}.
It has been suggested that this roughly universal LF is the result of 
fractal structure in turbulent gas \citep{EE97}.
Since the three M101 GHRs show different age and spatial distributions
of clusters, it is interesting to intercompare their cluster LFs and
see if they also follow the universal cluster LFs.

The LFs of clusters in NGC\,5461, NGC\,5462, and NGC\,5471 are 
presented in Figure~\ref{fig:lum_func}.
These LFs are constructed using raw $M_{\rm F547M}$ (without 
extinction correction).
No completeness correction to the LFs is needed because the
cutoff of our sample at the faint end, $m_{\rm F547M} \leq 21.3$,
is much brighter than the detection limit, $m_{\rm F547M} \sim 25.5$.
Note however that the two faintest bins, $M_{\rm F547M} = -8.0$ to 
$-$9.0, should be viewed with caution, as some of the ``clusters''
may be spurious as discussed in \S5.1.
The number of clusters in each GHR is modest, so we have also
constructed a combined cluster LF of the three GHRs, shown in the
bottom panel of Figure~\ref{fig:lum_func}.
We have carried out linear least-squares fits to the logarithmic 
LFs of clusters for the three GHRs individually and combined.
The logarithmic LFs of clusters in the individual GHRs appear to 
have different slopes; the best-fit slopes for NGC\,5461, 
NGC\,5462, and NGC\,5471 are $-1.5\pm0.3$, $-3.0\pm0.2$, 
and $-1.9\pm0.4$, respectively.
The logarithmic LF of clusters in all three GHRs has a best-fit 
slope of $-2.3\pm0.1$.

Compared with the universal cluster LFs, NGC\,5461 and NGC\,5471
are on the flatter side, and NGC\,5462 is on the steeper side.
The small difference between NGC\,5461 and NGC\,5471 is not
statistically significant, as each has only a small number of 
clusters and the numbers of clusters in the brightest bins
are only 1--2.
On the other hand, it may be statistically significant that
NGC\,5462 has a much steeper LF than NGC\,5461 and NGC\,5471,
or NGC\,5462 has a larger proportion of low-mass clusters.
It is possible that the slope of a cluster LF varies according
to the interstellar environment at the time when clusters 
were formed.
As discussed in \S5.3, the current molecular environments of 
NGC\,5461 and NGC\,5462 are quite different, with molecular
CO highly concentrated in NGC\,5461 and diffuse in NGC\,5462. 
If the current environments reflect the conditions when the
clusters were formed, the clusters in NGC\,5462 would have been
formed in a lower-pressure, lower-concentration interstellar
environment.
The association between a steep cluster LF and a low-pressure,
low-concentration star formation condition is consistent with
the previously suggested hypothesis that massive clusters are
formed in high-pressure, high-concentration molecular clouds.

The cluster LF for NGC\,5461, NGC\,5462, and NGC\,5471 combined 
has an $\alpha$ within the range of the universal value, 
$-$2 to $-$2.4.
If clusters in spiral or starburst galaxies are formed under 
conditions similar to NGC\,5461 and NGC\,5462 in random 
proportions, the cluster LFs should show a larger range of 
$\alpha$.
The scarcity of cluster LFs with $\alpha$ steeper than $-$2.5
might be an observational effect to some extent.
Surveys of clusters in galaxies preferentially detect the most 
luminous clusters that are likely formed in high-pressure 
environments and follow an LF similar to those of NGC\,5461 
and NGC\,5471; therefore, the cluster LF of a galaxy would be 
biased toward $\alpha = -2$.

\subsection{Evolutionary Aspects of the Clusters}

The cluster mass, age, and size distribution of a cluster
system may be used to investigate the dynamic evolution
of clusters.
Recent studies of rich clusters in the LMC have shown that the 
spread in core radius increases with cluster age, suggesting that 
all clusters were formed with small core radii but subsequently some 
experienced core expansion while others did not \citep{EFL89,MG03}. 
It would be interesting to examine the clusters in M101 GHRs to 
see whether they follow the same core radius-age relation.

The mass, age, and core radius of the LMC clusters have been 
derived by \citet{MG03}, using the \citet{FRV97} population 
synthesis code, the \citet{KTG93} IMF slope, and a stellar
mass range of 0.1--120 M$_\odot$.
To compare the M101 clusters with the LMC clusters, we have
followed the \citet{MG03} method and re-estimated the masses 
of M101 clusters, using the same code, IMF slope, and stellar 
mass range.
The new cluster mass estimates are $\sim$ 2--3 times as high as 
the cluster masses estimated earlier in this paper, owing to the
addition of stars in the mass range of 0.1--1 M$_\odot$.
The M101 clusters are not sufficiently resolved for measurements
of their core radii ($r_{\rm c}$); therefore, we have adopted the 
relation $r_{\rm c} \sim 0.35 R_{\rm eff}$ derived from a King 
profile with a concentration parameter of $c=1.5$, a median value
for LMC clusters \citep{EFF87}.
The core radii of M101 clusters thus estimated are 0.25--1 pc,
and may be uncertain by up to a factor of 2, if its concentration
parameter spans the same range as that in the LMC clusters.

To compare the M101 clusters with the LMC clusters, we
present a 3-D diagram of cluster mass, age, and core radius
in Figure~\ref{fig:3d-1}.
The data of the LMC clusters are adopted from \citet{MG03}.
The M101 clusters are plotted in open rhombuses, while the LMC 
clusters are plotted in filled ellipses, with the R136 cluster
in a larger ellipse for easy identification.
The M101 cluster masses were estimated using 7-pc-radius
apertures, so we repeated Mackey \& Gilmore's derivation of
the mass of R136 using this larger aperture and obtained a
mass that is 40\% higher, shown as a large open ellipse in 
Figure~\ref{fig:3d-1}.
While the LMC clusters span a large age range and show an
increasing spread in core radii with the cluster age,
the M101 clusters we analyzed are all young and small, 
sharing a similar parameter space with the R136 cluster.
Among the small number of M101 clusters with size measurements,
the younger clusters are generally smaller, but the accuracy of
the size measurements is too limited by the linear resolution 
for definitive conclusions.

Finally, we discuss the disruption time of our M101 clusters
and investigate whether dynamical evolution has caused a
significant mass change in these clusters.
The disruption time can be assessed empirically from the
break in slopes of the logarithmic age distribution of 
clusters \citep{BL03}, or from comparisons with $N$-body 
simulations \citep{BM03}.
Recent studies using the empirical method have shown that 
the disruption time for a 10$^4$ M$_\odot$ cluster
varies greatly among galaxies, from 10$^7$ to 10$^{10}$ yr,
with the shortest being $\sim$30--40 Myr in M82 and at 
1--3 kpc from the nucleus of M51 \citep{BL03,dG03}.
Our M101 clusters in GHRs are apparently younger 
than these disruption timescales.
Thus we determine the disruption time, $T_{\rm dis}$, following
the simulations of \citet{BM03}:
\begin{equation}
\frac{T_{\rm dis}}{\rm Myr} = \beta~\left[\frac{N}{{\rm ln}(\gamma N)} 
\right]^{x} \frac{R_{\rm G}}{\rm kpc} \left( \frac{V_{\rm G}}{220~
{\rm km~s}^{-1}} \right) ^{-1}
\end{equation}
where $R_{\rm G}$ is the galactocentric radius, $V_{\rm G}$ is 
the circular velocity in a galaxy, $N$ is the number of stars in
a cluster, $\beta \sim$ 1--2, $\gamma$ = 0.02, and $x \sim$ 0.8.
For a cluster with a Salpeter mass function and a stellar
mass range of 1--100 M$_\odot$, the average mass of a star is
3.09 M$_\odot$ and the number of stars is  
$N$ = cluster mass/(3.09 M$_\odot$).
NGC\,5461 and NGC\,5462 have $R_{\rm G} \sim 10$ kpc and
$V_{\rm G}$ = 185 km~s$^{-1}$, and NGC\,5471 has $R_{\rm G} 
\sim 25$ kpc and $V_{\rm G}$ = 195 km~s$^{-1}$ \citep{RR73}.
A 10$^4$ M$_\odot$ cluster in NGC\,5461/NGC\,5462 or NGC\,5471
would have disruption times of 2.4--4.8 $\times10^9$ and 
5.8--11.6 $\times10^9$ yr, respectively.
The real disruption time must be shorter because the above estimates
do not take into account processes that are important in disrupting
clusters in GHRs, i.e., interactions with other clusters and with
giant molecular clouds.
In cases where massive clusters are concentrated in a small
volume, such as the core of NGC\,5461, cluster merger is a 
more important dynamic process and operates in a much shorter
timescale than tidal disruption \citep{BBV03}.
Future simulations of dynamical evolution of clusters in GHRs
using realistic conditions are needed.

\section{Summary}

GHRs contains high concentrations of massive stars; thus, they 
provide an excellent laboratory to study modes of massive star 
formation and possible sites to form globular clusters.
We have selected three very luminous but morphologically 
different GHRs in M101, NGC\,5461, NGC\,5462, and NGC\,5471,
to determine their cluster content in order to understand
cluster formation in different environments.
We have obtained {\it HST} WFPC2 images of these GHRs with the 
F547M and F675W continuum filters and the F656N H$\alpha$ filter.
The continuum images are used to identify cluster candidates
in each GHR and to carry out photometric measurements, and the
H$\alpha$ images are used to examine the distribution of
interstellar gas and to determine the ionizing flux requirement.

We have used the Starburst99 \citep{Le99} and BC03 \citep{BC03} 
population synthesis models to compute the colors and magnitudes
of clusters of different ages and masses.
The colors of a cluster are dependent on its age; however,
our two continuum passbands are not blue enough to be
sensitive to young massive stars for age determination.
Therefore, we use the distribution of ionized interstellar gas 
to estimate the approximate cluster ages, then compare the
measured colors and magnitudes of cluster candidates to 
the synthetic evolutionary tracks to determine their masses.
To avoid confusion by luminous single supergiants, only
cluster candidates more luminous than $M_{\rm F547M} = -9.0$
are analyzed for their masses.

NGC\,5461 is dominated by a very luminous core, and has been
suggested as a likely host of SSCs \citep{KC88,LP01}.
Our observations show that the core of NGC\,5461 contains three 
R136-class clusters superposed on a bright stellar background
in a small region $\sim$32 pc across.
It is possible that the three R136-class clusters will dynamically 
interact and merge into an SSC.
If NGC\,5461 were at a distance $\gtrsim$ 20 Mpc, the clusters at
its core would appear as a single cluster, and the total light
would be $M_{\rm F547M} = -13.0$, corresponding to a mass of 
$\sim 10^5$ M$_\odot$, reaching those of SSCs.
It is possible that some of the previously reported SSCs at large
distances are actually made up by tight groups of R136-class 
clusters similar to those in NGC\,5461.

NGC\,5462 consists of numerous loosely-distributed \hii\ 
regions that are individually much fainter than 30 Dor.
Its clusters also show a loose distribution across the GHR.
NGC\,5462 has the largest number of clusters among the three 
GHRs studied, but most of the clusters are older than 10 Myr 
and fainter than $M_{\rm F547M}=-9.0$.

NGC\,5471 contains multiple bright \hii\ regions, some of which
are comparable to 30 Dor.
A large number of cluster candidates are identified in NGC\,5471; 
the majority of the clusters are fainter than $M_{\rm F547M}=-9.0$
and they are in bright \hii\ regions.
The mass determination for clusters in NGC\,5471 is problematic
because the observed cluster colors are bluer than those 
spanned by the synthetic cluster evolutionary tracks for 
$Z = 0.2 Z_\odot$, possibly as a result of uncertainties
in stellar evolution models at low metallicities.
The cluster masses are thus estimated from the magnitudes
alone and may be subject to large errors.

The most massive clusters in the three GHRs are in the
mass range of $\sim$ 1--3 $\times 10^4$ M$_\odot$, similar 
to R136.
Two clusters in NGC\,5471 might be more massive as they
are not surrounded by \hii\ regions and are each as luminous 
as R136; these two may be the most massive clusters in the 
three GHRs studied.
No SSCs are present in any of the three GHRs.
We have also estimated the sizes of some clusters on their PC 
images, using the routine developed by \citet{La99}.
The effective radii of these clusters are in the range of 
0.7--2.9 pc, $\sim$ 1--3 times that of R136 \citep{MG03}.

To understand the makeup of the faint cluster candidates,
we have simulated WFPC2 images of 30 Dor and NGC\,604 at
the distance of M101.
We find that single supergiants similar to R131 and OB 
associations as those in NGC\,604 may contribute to the 
population of faint clusters ($M_{\rm F547M} > -9.0$) in
distant galaxies, while clusters similar to Hodge 301 
or OB associations similar to LH\,99 will be too faint or
too extended to be identified as clusters even in M101.

The three M101 GHRs show different cluster LFs.  
The cluster LFs of spiral galaxies can be described by
a power-law with the exponent $\alpha$ in the range of 
$-$2.0 to $-$2.4 \citep{La02}.
We find that the cluster LFs of NGC\,5461 and NGC\,5471 
are on the flatter side of the range, but the number of
clusters is small in each GHR.
NGC\,5462 has the largest number of clusters and its cluster
LF is significantly steeper, with $\alpha = -3.0\pm0.2$.
It is possible that the clusters in NGC\,5462 were formed 
in a low-pressure, low-concentration interstellar environment.
The combined cluster LF of the three GHRs has an $\alpha$
of $-2.3\pm0.1$, well within the range for those of spiral
galaxies.
The universality of cluster LFs may be a statistical
result from a cluster population with an observational
bias toward the most luminous clusters.

The distribution of molecular clouds is concentrated
in NGC\,5461 and diffuse in NGC\,5462, similar to the
spatial distribution of their clusters.
The diffuse interstellar environment and the larger proportion
of low-mass clusters (steep cluster LF) of NGC\,5462 qualitatively
support the hypothesis that massive clusters are formed in 
high-pressure, high-concentration interstellar medium.

We have estimated the fraction of massive stars formed in
clusters using (1) clusters' contribution to the total
stellar continuum, and (2) comparison between the ionizing
flux expected from the clusters and the ionizing flux
required by the associated \hii\ region.
Both methods show that $\lesssim$ 50\% of massive 
stars are formed in R136-class clusters.
Consequently, the \ha\ luminosity of an \hii\ region does
not provide a sufficient diagnostic for the existence of SSCs.

\acknowledgments
This research is supported by grants STScI GO-6829.01-95A
and GO-9934.01-A, and uses observations conducted for program 
GO-6829 with the NASA/ESA {\it HST}, obtained at the Space 
Telescope Science Institute, which is operated by the 
Association of Universities for Research in Astronomy, Inc., 
under NASA contract NAS 5-26555. 
K.\ E.\ J. gratefully acknowledges support for this work provided by 
NSF through and Astronomy and Astrophysics Postdoctoral Fellowship. 
We have used the Digital Sky Survey produced at the Space 
Telescope Science Institute under US government grant NAG W-2166.
We thank the anonymous referee for prompt reviewing 
and making useful suggestions to improve this paper.
We also thank R.\ C.\ Smith and C.\ Aguilera for providing
MCELS green-band images of 30 Dor, F.\ Schweizer for suggesting 
references on core radius-mass relation, G.\ Bruzual and S.\ 
Charlot for providing the population synthesis code, and 
R.\ Gruendl, R.\ Williams, and B.\ Dunne for reading the 
manuscript.

\clearpage

\begin{figure}
\caption{The POSS-II red image of M101 from the Digitized Sky Survey.  
 The three luminous GHRs studied in this paper,  NGC\,5461, NGC\,5462, 
 and NGC\,5471, are marked.
 \label{fig:m101}}
\end{figure}

\begin{figure}
\caption{Color composite of {\it HST} WFPC2 images of NGC\,5461,
  with F547M in blue, F675W in green, and \ha\ in red.
 North is up and east to the left.  The field-of-view is 
 $45\arcsec\times45\arcsec$, or 1.6 kpc $\times$ 1.6 kpc.
 \label{color1}} 
\end{figure}

\begin{figure}
\caption{Color composite of {\it HST} WFPC2 images of NGC\,5462,
 with F547M in blue, F675W in green, and \ha\ in red.
 North is up and east to the left.  The field-of-view is 
 $56\arcsec\times56\arcsec$, or 2.0 kpc $\times$ 2.0 kpc. 
 \label{color2}} 
\end{figure}

\begin{figure}
\caption{Color composite of {\it HST} WFPC2 images of NGC\,5471,
  with F547M in blue, F675W in green, and \ha\ in red.
 North is up and east to the left.  The field-of-view is 
 $51\arcsec\times51\arcsec$, or 1.8 kpc $\times$ 1.8 kpc. 
 \label{color3}} 
\end{figure}

\begin{figure}
\caption{{\it HST} WFPC2 images of the main body of NGC\,5461 in  (a) F547M, 
 (b) F675W, (c) \ha, (d) \ha-subtracted F675W, (e) continuum-subtracted \ha, 
 and (f) F547M bands.  
 Note that this field-of-view is smaller than that shown in Figure~2.  
 The \ha\ images (c) and (e) are presented in different stretches to show 
 bright and faint features, and similarly the F547M images (a) and (f)
 to show bright and faint stars/clusters. 
 The two brightest \hii\ regions from \citet{Ho90} are marked in (e),
 and cluster candidates are marked in (f). \label{fig:n5461}}
\end{figure}

\begin{figure}
\caption{{\it HST} WFPC2 images of the main body of NGC\,5462,
 displayed in a format identical to that of Figure~5.
 Note that this field-of-view is smaller than that shown in Figure~3.
 The three brightest \hii\ regions from \citet{Ho90} are marked in (e),
 and cluster candidates are marked in (f). \label{fig:n5462}}
\end{figure}

\begin{figure}
\caption{{\it HST} WFPC2 images of NGC\,5471, displayed in a format 
 identical to that of Figure~5.
 Note that this field-of-view is smaller than that shown in Figure~4.
 The five brightest components from \citet{Sk85} are marked in (e),
 and cluster candidates are marked in (f). \label{fig:n5471}}
\end{figure}


\begin{figure}
\caption{$M_{\rm F547M}$ versus ($M_{\rm F547M}-M_{\rm F675W'}$) diagram 
 of cluster candidates in NGC\,5461.
 Observations of the clusters are plotted in filled circles.
 Additional circles are draw around clusters that are embedded
 in bright \hii\ regions.
 Evolutionary tracks generated from Starburst99 and BC03 for a Salpeter 
 initial mass function and a metallicity of $Z = 1~Z_\odot$ are plotted 
 in solid and dotted curves, respectively.
 Ages in Myr are marked along the evolutionary tracks.
 To avoid crowding, the Starburst99 evolutionary track is shown
 for a cluster mass of $5\times10^4$ M$_\odot$ and the BC03 track for
 $1\times10^4$ M$_\odot$.
 The reddening vectors of associated \hii\ regions are plotted in dashed 
 arrows, and the possible dereddening vectors are marked with solid 
 arrows for clusters brighter than $M_{\rm F547M} = -9.0$. 
 The R136 cluster is plotted as a reference point: reddened R136
 in a filled triangle and dereddened R136 in an open triangle.
 \label{fig:cmd_5461}}
\end{figure}

\begin{figure}
\caption{$M_{\rm F547M}$ versus ($M_{\rm F547M}-M_{\rm F675W'}$) 
 diagram of cluster candidates in NGC\,5462.  Symbols are the same 
 as in Figure~8 with the addition of a dashed circle for the cluster 
 candidate in an interstellar shell.
 The evolutionary tracks are the same as in Figure~8.
 \label{fig:cmd_5462}}
\end{figure}

\begin{figure}
\caption{$M_{\rm F547M}$ versus ($M_{\rm F547M}-M_{\rm F675W'}$) 
 diagram of cluster candidates in NGC\,5471.  Symbols are the same 
 as in Figure~8.  The evolutionary tracks are generated from population 
 synthesis models using the same parameters as those generated for
 NGC\,5461 and NGC\,5462, but with a metallicity of $Z = 0.2~Z_\odot$. 
 \label{fig:cmd_5471}}
\end{figure}

\begin{figure}
\caption{(a) {\it HST} WFPC2 PC image of the clusters in the core 
 of NGC\,5461 (i.e., H\,1105) in the F547M band.  The cluster numbers
 given in Figure 5 are again marked.  (b) Binned 
 {\it HST} WFPC2 WFC image in the F547M band for the same region to 
 simulate a WFC image at 20 Mpc.  At such a large distance, the
 three clusters at the core of NGC\,5461 are no longer distinguishable
 from a single super-star cluster. \label{fig:5461core}}
\end{figure}

\begin{figure}
\caption{{\it HST} WFPC2 PC images of the brightest clusters in 
 NGC\,5462 and NGC\,5471 for which cluster sizes were measured: 
 (a) NGC\,5462-6, (b) NGC\,5471-4 \& -5, (c) NGC\,5471-9, and 
 (d) NGC\,5471-16.  The pixel size of a PC image is 0\farcs0455, 
 corresponding to a linear size of 1.6 pc in M101.  The field-of-view 
 of each image is 2\farcs4 $\times$ 2\farcs4. \label{fig:clusters}}
\end{figure}

\begin{figure}
\caption{(a) Binned MCELS green band image of 30 Dor and (b)
 binned {\it HST} WFPC2 F547M images of NGC\,604 to simulate 
 WFPC2 images of 30 Dor and NGC\,604 at the distance of M101. 
 (c)--(e) {\it HST} WFPC2 F547M images of M101 GHRs NGC\,5461, 
 NGC\,5462, and NGC\,5471.  The cluster candidates with $M_{\rm F547M}
 > -9.0$ are marked.  All images have the same linear field-of-view,
 350 pc $\times$ 350 pc. The four F547M images are displayed 
 with the same intensity scale, while the green band image of 30 
 Dor is displayed with an intensity scale matching that of the 
 F547M images as much as possible. \label{fig:ghrs}}
\end{figure}

\begin{figure}
\caption{{\it HST} WFPC2 images of four regions with R136-class 
 clusters coexistent with bright diffuse stellar background in 
 the F547M band (upper panels) and \ha\  band (bottom panels): 
(a) NGC\,5461-1, (b) NGC\,5461-8, 9, 10, 
 (c) NGC\,5462-18, and (d) NGC\,5471-16.
 Circles in the F547M images mark the apertures used to measure the 
 total stellar continuum emission of each region, and ellipses 
 in the \ha\ images mark the apertures used to measure the \ha\ 
 fluxes of the associated \hii\ region.
 The field-of-view of each image is $8'' \times 8''$.
 \label{fig:4regions}}
\end{figure}


\begin{figure}
\caption{$M_{\rm F547M}$ luminosity functions (LFs) of candidate 
 clusters in NGC\,5461, NGC\,5462, NGC\,5471, and of the combined 
 sample of all three GHRs. Solid lines are cluster LFs and dotted 
 lines are power-law fits to these LFs.  The best-fit $\alpha$ value 
 and the number of cluster candidates are labeled in the upper
 left corner of each panel. 
 \label{fig:lum_func}}
\end{figure}

\begin{figure}
\caption{A 3-D diagram of age, core radius, and mass of clusters in
  the LMC and M101.  The M101 
 clusters are shown as open rhombuses, while the LMC clusters 
 \citep{MG03} are shown as filled ellipses.  The R136 cluster is 
 shown as a larger filled ellipse and open ellipse for the mass
 estimates from \citet{MG03} and this study, respectively. 
 \label{fig:3d-1}} 
\end{figure}

\clearpage

\begin{deluxetable}{lcccc}
\tablecaption{Properties of the Three Luminous GHRs in M101 and 
 30 Dor in the LMC  \label{tbl:GHRs}}
\tablewidth{0pt}
\tablehead{
\colhead{} & \colhead{NGC\,5461} & \colhead{NGC\,5462} & 
\colhead{NGC\,5471} & \colhead{30 Dor}}

\startdata

Angular Size & $40'' \times 25''$ & $48'' \times 33''$ & $17'' \times 17''$ 
	& $20' \times 20'$ \\
Linear Size\tablenotemark{a} (pc)  & $1400\times875$~~ & $1680\times1150$ 
	& $600 \times 600$ & $290 \times 290$ \\ 
$L_{\rm H\alpha}$\tablenotemark{b} (ergs~s$^{-1})$ & ~$2.7\times10^{40}$ 
	& ~\,$1.3\times10^{40}$ & ~\,$2.2\times10^{40}$ 
	& ~\,$3.9\times10^{39}$\\
Location& in spiral arm & in spiral arm & outlier & above one end of  \\
	&		&		&	  & the LMC bar	      \\
H$\alpha$ Morphology &one dominant core & weak cores with 
	& multiple cores  & one core with\\
	& with filaments \&  & long filaments \&
	& with filaments	& bright loops \&\\
	& small cores around  & loops extending out
	& around		& filaments around\\

\enddata
\tablenotetext{a}{We adopted the distances of 7.2 Mpc to M101 \citep{St98}
 and 50 kpc to the LMC \citep{Fe99}.}
\tablenotetext{b}{The $L_{\rm H\alpha}$ of NGC\,5461, NGC\,5462, and
 NGC\,5471 are measured using {\it HST} WFPC2 \ha\ images in this
 study, and the $L_{\rm H\alpha}$ of 30 Dor is adopted from \citet{KH86}.
 Note that these $L_{\rm H\alpha}$ are not corrected for extinction.}
\end{deluxetable}


\begin{deluxetable}{crccl}
\tablecaption{Table of Observations \label{tbl:log}}
\tablewidth{0pt}
\tablehead{
\colhead{} & \colhead{Obs. Date} & \colhead{} & 
\colhead{} & \colhead{}\\
\colhead{Object} & \colhead{(y/m/d)} & \colhead{Filter} & 
\colhead{Camera} & \colhead{Exp. Time}}

\startdata

NGC\,5461& 1999/03/24 & F547M & WF2 & 600s $\times2$, 100s $\times2$, 
  ~\,20s $\times1$\\
	&             & F547M & PC1 & ~\,20s $\times2$ \\
	&  	      & F675W & WF2 & 400s $\times2$, ~\,50s $\times2$,
  ~\,10s $\times1$\\
	& 1999/03/23  & F656N & WF2 & 600s $\times2$, 160s $\times1$   \\
NGC\,5462 & 2000/02/01& F547M & WF2 & 600s $\times2$, 100s $\times2$,
  ~\,20s $\times1$\\
	&             &	F547M & PC1 & ~\,20s $\times2$ \\
	& 	      & F675W & WF2 & 400s $\times2$, ~\,50s $\times2$,
  ~\,10s $\times1$\\
	& 	      & F656N & WF2 & 600s $\times2$, 160s $\times1$  \\
NGC\,5471 & 1997/11/01& F547M & WF3 & 600s $\times2$, 100s $\times2$,
  ~\,20s $\times1$\\
	&	      & F547M & PC1 & ~\,20s $\times2$ \\
	&	      & F675W & WF3 & 400s $\times2$, ~\,50s $\times2$,
  ~\,10s $\times1$\\
	&	      & F656N & WF3 & 600s $\times2$, 180s $\times1$  \\

\enddata
\end{deluxetable}


\begin{deluxetable}{rrrccc}
\tablecaption{Photometry of Candidate Clusters in NGC\,5461 
 \label{tbl:photos-1}}
\tablewidth{0pt}
\tablehead{
\multicolumn{1}{c}{ID} &  \colhead{$\alpha_{\rm J2000}$} &   
\colhead{$\delta_{\rm J2000}$}&
\colhead{$M_{\rm F547M}$} & \colhead{$M_{\rm F547M}-M_{\rm F675W}$} & 
\colhead{$M_{\rm F547M}-M_{\rm F675W'}$} }

\startdata

 1 &14 03 39.84 &54 18 56.2 & $-9.35\pm$0.01& 0.45$\pm$0.05&  0.23$\pm$0.02 \\
 2 &14 03 39.91 &54 18 56.2 & $-8.19\pm$0.05& 1.26$\pm$0.06&  0.55$\pm$0.06 \\
 3 &14 03 40.52 &54 18 58.8 & $-8.97\pm$0.01& 0.17$\pm$0.03&  0.16$\pm$0.02 \\
 4 &14 03 40.54 &54 18 59.2 & $-8.11\pm$0.03& 0.14$\pm$0.08&  0.11$\pm$0.04 \\
 5 &14 03 40.98 &54 19 02.1 & $-8.25\pm$0.02& 0.11$\pm$0.06&  0.09$\pm$0.03 \\
 6 &14 03 41.15 &54 19 04.5 & $-9.61\pm$0.03& 0.21$\pm$0.09&  0.10$\pm$0.04 \\
 7 &14 03 41.22 &54 18 57.0 & $-8.01\pm$0.02& 0.17$\pm$0.02&  0.13$\pm$0.02 \\
 8\tablenotemark{a} &14 03 41.36 &54 19 03.7 &
   $-10.40\pm$0.10~\,&  0.59$\pm$0.13 &  0.36$\pm$0.10 \\
			 &&&$-10.19\pm$0.06~\,&&\\
 9\tablenotemark{a} &14 03 41.40 &54 19 03.8 &
   $-10.17\pm$0.12~\,&  0.39$\pm$0.16 &  0.24$\pm$0.14 \\
			 &&&$-9.75\pm$0.11&&\\
10\tablenotemark{a} &14 03 41.42 &54 19 04.0 &
   $-10.81\pm$0.06~\,&  0.27$\pm$0.09 &  0.22$\pm$0.07 \\
			 &&&$-10.57\pm$0.06~\,&&\\
11 &14 03 41.58 &54 19 04.0 & $-9.27\pm$0.05& 1.37$\pm$0.05&  0.58$\pm$0.05 \\
12 &14 03 41.58 &54 19 07.8 & $-8.37\pm$0.01& 0.38$\pm$0.03&  0.38$\pm$0.02 \\

\enddata
\tablenotetext{a}{The photometry is given in two rows, with the 
first row measured with the WFC images and the second row measured with 
the PC images.}
\end{deluxetable}


\begin{deluxetable}{rrrrcc}
\tablecaption{Photometry of Candidate Clusters in NGC\,5462 \label{tbl:photos-2}}
\tablewidth{0pt}
\tablehead{
\multicolumn{1}{c}{ID} &  \colhead{$\alpha_{\rm J2000}$} &   
\colhead{$\delta_{\rm J2000}$}&
\colhead{$M_{\rm F547M}$} & \colhead{$M_{\rm F547M}-M_{\rm F675W}$} & 
\colhead{$M_{\rm F547M}-M_{\rm F675W'}$} }

\startdata

 1 &14 03 51.80 &54 21 52.6 & $-8.83\pm$0.04& 0.32$\pm$0.05&  0.16$\pm$0.04 \\
 2 &14 03 51.83 &54 21 46.2 & $-8.08\pm$0.01& 0.31$\pm$0.02&  0.27$\pm$0.02 \\
 3 &14 03 51.98 &54 21 46.7 & $-8.64\pm$0.01& 0.36$\pm$0.02&  0.33$\pm$0.02 \\
 4 &14 03 52.07 &54 21 49.1 & $-8.32\pm$0.01& 0.46$\pm$0.01&  0.43$\pm$0.01 \\
 5 &14 03 52.38 &54 21 49.5 & $-8.33\pm$0.02& 0.20$\pm$0.03&  0.15$\pm$0.03 \\
 6 &14 03 52.41 &54 21 49.1 & $-9.56\pm$0.01& 0.43$\pm$0.01&  0.40$\pm$0.01 \\
 7 &14 03 52.42 &54 21 49.4 & $-8.16\pm$0.05& 0.47$\pm$0.05&  0.44$\pm$0.05 \\
 8 &14 03 52.43 &54 21 50.0 & $-8.86\pm$0.02& 0.30$\pm$0.02&  0.28$\pm$0.02 \\
 9 &14 03 52.84 &54 21 54.5 & $-8.05\pm$0.02& 0.14$\pm$0.03&  0.12$\pm$0.03 \\
10 &14 03 52.86 &54 21 59.3 & $-8.13\pm$0.02& 0.63$\pm$0.02&  0.59$\pm$0.02 \\
11 &14 03 52.95 &54 21 54.2 & $-8.55\pm$0.01& 0.65$\pm$0.02&  0.52$\pm$0.02 \\
12 &14 03 52.96 &54 22 06.4 & $-8.75\pm$0.03& 1.41$\pm$0.03&  0.37$\pm$0.03 \\
13 &14 03 53.01 &54 22 00.6 & $-8.44\pm$0.02& 0.51$\pm$0.02&  0.45$\pm$0.02 \\
14 &14 03 53.03 &54 21 56.1 & $-8.10\pm$0.02& 0.23$\pm$0.03&  0.20$\pm$0.03 \\
15 &14 03 53.35 &54 22 00.5 & $-8.87\pm$0.01& 0.20$\pm$0.01&  0.17$\pm$0.01 \\
16 &14 03 53.41 &54 21 59.1 & $-8.61\pm$0.01& 0.41$\pm$0.02&  0.39$\pm$0.02 \\
17 &14 03 53.58 &54 22 04.0 & $-8.03\pm$0.02& 0.44$\pm$0.03&  0.42$\pm$0.02 \\
18 &14 03 53.78 &54 22 11.1 & $-9.03\pm$0.02& 0.28$\pm$0.08&  0.03$\pm$0.03 \\
19 &14 03 53.98 &54 21 56.1 & $-8.24\pm$0.01& 0.50$\pm$0.02&  0.48$\pm$0.02 \\
20 &14 03 54.00 &54 22 07.9 & $-8.35\pm$0.02& 0.19$\pm$0.03&  0.18$\pm$0.03 \\
21 &14 03 54.10 &54 22 02.5 & $-8.30\pm$0.01& 0.54$\pm$0.02&  0.41$\pm$0.02 \\
22 &14 03 54.18 &54 22 06.6 & $-8.24\pm$0.02& 0.33$\pm$0.03&  0.31$\pm$0.03 \\
23 &14 03 54.19 &54 22 11.2 & $-9.17\pm$0.01& 0.55$\pm$0.01&  0.51$\pm$0.01 \\
24 &14 03 54.32 &54 22 09.1 & $-8.09\pm$0.05& 0.10$\pm$0.06&  0.07$\pm$0.06 \\
25 &14 03 54.74 &54 21 53.7 & $-8.20\pm$0.01& 0.27$\pm$0.01&  0.24$\pm$0.01 \\

\enddata

\end{deluxetable}


\begin{deluxetable}{rrrccc}
\tablecaption{Photometry of Candidate Clusters in NGC\,5471 \label{tbl:photos-3}}
\tablewidth{0pt}
\tablehead{
\multicolumn{1}{c}{ID} &  \colhead{$\alpha_{\rm J2000}$} &   
\colhead{$\delta_{\rm J2000}$}&
\colhead{$M_{\rm F547M}$} & \colhead{$M_{\rm F547M}-M_{\rm F675W}$} & 
\colhead{$M_{\rm F547M}-M_{\rm F675W'}$} }

\startdata

 1\tablenotemark{a} &14 04 28.64 &54 23 51.9 &
  $-8.81\pm$0.09&  1.24$\pm$0.12&  0.41$\pm$0.11 \\
			  &&&$-8.23\pm$0.12&&\\
 2\tablenotemark{a} &14 04 28.64 &54 23 52.1 &
  $-9.31\pm$0.08&  1.46$\pm$0.10&  0.76$\pm$0.10 \\
			  &&& $-9.05\pm$0.09&&\\
 3 &14 04 28.86 &54 23 48.2 & $-9.08\pm$0.03& $-0.18\pm$0.06~~\,& 
  $-0.13\pm$0.04~~\, \\
 4 &14 04 28.88 &54 23 47.8 & $-9.86\pm$0.02& 0.15$\pm$0.03& 0.12$\pm$0.03 \\
 5 &14 04 28.90 &54 23 48.4 &$-10.27\pm$0.01~\,& $-0.02\pm$0.02~~\,& 
  $-0.07\pm$0.02~~\, \\
 6 &14 04 29.10 &54 23 41.7 & $-8.59\pm$0.04& 0.34$\pm$0.09& $-0.09\pm$0.05~~\, \\
 7 &14 04 29.15 &54 23 45.9 & $-9.00\pm$0.06& 0.04$\pm$0.09& $-0.11\pm$0.07~~\, \\
 8 &14 04 29.17 &54 23 45.4 & $-8.87\pm$0.06& 1.12$\pm$0.08&  0.14$\pm$0.07 \\
 9 &14 04 29.27 &54 23 51.2 & $-9.38\pm$0.01& 0.26$\pm$0.03&  0.19$\pm$0.01 \\
10 &14 04 29.29 &54 23 52.4 & $-8.71\pm$0.08& 0.99$\pm$0.11&  0.14$\pm$0.13 \\
11 &14 04 29.29 &54 23 52.9 & $-8.21\pm$0.05& 0.86$\pm$0.09&  0.22$\pm$0.08 \\
12 &14 04 29.33 &54 23 47.2 & $-9.06\pm$0.02& 0.37$\pm$0.04&  0.19$\pm$0.03 \\
13 &14 04 29.37 &54 23 46.5 & $-8.47\pm$0.04& 0.34$\pm$0.04& $-0.04\pm$0.04~~\, \\
14 &14 04 29.38 &54 23 46.2 & $-8.08\pm$0.06& 0.72$\pm$0.08&  0.35$\pm$0.09 \\
15 &14 04 29.39 &54 23 51.8 & $-8.02\pm$0.02& 0.54$\pm$0.14&  0.02$\pm$0.09 \\
16 &14 04 29.47 &54 23 46.4 &$-10.06\pm$0.06~\,&0.16$\pm$0.13& $-0.13\pm$0.06~~\, \\
17 &14 04 29.53 &54 23 46.1 & $-8.68\pm$0.14& 1.47$\pm$0.15&  0.25$\pm$0.15 \\
18 &14 04 29.54 &54 23 45.8 & $-8.40\pm$0.11& 1.38$\pm$0.13&  0.52$\pm$0.12 \\
19 &14 04 29.56 &54 23 47.5 & $-8.97\pm$0.02& 0.44$\pm$0.07&  0.21$\pm$0.03 \\

\enddata
\tablenotetext{a}{The photometry is given in two rows, with 
the first row measured with the WFC images and the second row measured 
with the PC images.}
\end{deluxetable}


\begin{deluxetable}{lccccc}
\tablecaption{Physical Properties of Massive Clusters in M101 GHRs 
\label{tbl:mass}}
\tablewidth{0pt}
\tablehead{
\colhead{} & \colhead{Age} & \colhead{Starburst99 Mass} & 
 \colhead{BC03 Mass} & \colhead{R$_{\rm eff}$} &  \colhead{}  \\
\colhead{Cluster ID} & \colhead{(Myr)} & \colhead{($\times 10^4 M_\odot$)}
 & \colhead{($\times 10^4 M_\odot$)} & \colhead{(pc)} & \colhead{Remarks} }

\startdata

NGC\,5461-1    & $<$ ~\,5 &~~2$\pm$0.5& ~~2$\pm$0.5 & ... & \\
NGC\,5461-6    & $<$ ~\,5 &~~1$\pm$0.5 & $\lesssim$ 1 & $0.8\pm0.2$ & \\
NGC\,5461-8    & $<$ ~\,5 &1.5$\pm$0.5& 1.5$\pm$0.5 & $0.7\pm0.2$ &  \\
NGC\,5461-9    & $<$ ~\,5 &1.5$\pm$0.5& ~~1$\pm$0.5 & ... &  \\
NGC\,5461-10   & $<$ ~\,5 &3$\pm$1& 2$\pm$1 & $2.1\pm0.1$ & asymmetric 
 morphology\\
NGC\,5461-11   & $<$ ~\,5 &~~~$\sim$ 0.5& ~~~$\lesssim$ 0.5 & ... & \\
\cline{1-6}
NGC\,5462-6    & $>$ 10	&$\sim$ 2 & 1.5--2 & $2.3\pm0.3$ & asymmetric 
 morphology\\
NGC\,5462-18   & $<$ ~\,5 &$\lesssim$ 1	& 0.5--1 & ... &  \\
NGC\,5462-23   & $>$ 10	&$\sim$ 2	& 1.5--2 & ... &  \\
\cline{1-6}
NGC\,5471-2    & $<$ ~\,5 &~~$\sim$ 0.2 & ~~$\sim$ 0.2 & ...&  \\
NGC\,5471-3    & $>$ 10   &~~$\sim$ 0.5 & ~~$\sim$ 0.5 & ...&  \\
NGC\,5471-4    & $>$ 10   &$\gtrsim$ 1	& $\gtrsim$ 1 &
  $2.9\pm0.3$ & asymmetric morphology \\
NGC\,5471-5    & $>$ 10   &$\sim$ 2 & $\sim$ 2 & 
  $1.4\pm0.1$ & asymmetric morphology \\
NGC\,5471-7    & $<$ ~\,5 &~~$\sim$ 0.5 &~~$\sim$ 0.5 & ... &  \\
NGC\,5471-9    & $>$ 10   &$\sim 1$ & $\sim$ 1 & $0.2\pm0.1$ & 
 probably a star \\
NGC\,5471-12   & $<$ ~\,5 &~~$\sim$ 0.5 &~~$\sim$ 0.5 & ... &  \\
NGC\,5471-16   & $<$ ~\,5 &1.5$\pm$0.5~ & 1.5$\pm$0.5~ & $1.1\pm0.1$ & \\

\enddata
\end{deluxetable}


\begin{deluxetable}{ccccccccc}
\tabletypesize{\scriptsize}
\tablecaption{Fractional Contribution of Clusters to Total Stellar Light
 \& Ionizing Luminosity}
\tablewidth{0pt}
\tablehead{
\colhead{} & \colhead{} & \colhead{Continuum} & \colhead{} & \colhead{} &
 \colhead{H$\alpha$} & \colhead{} & \colhead{} \\ 
\colhead{} & \colhead{} & \colhead{Aperture} &
 \colhead{$\frac{L_{\rm cluster}}{L_{\rm total}}$} & 
 \colhead{$\frac{L_{\rm cluster}}{L_{\rm total}}$} &
 \colhead{Aperture} &
 \colhead{$Q_{\rm H II}$} & \colhead{$Q_{\rm cluster}$\tablenotemark{a}} &
 \colhead{} \\ 
\colhead{Region} & \colhead{Cluster} & \colhead{Diameter} & 
 \colhead{in F547M} & \colhead{in F675W$'$} &  \colhead{Diameter} & 
 \colhead{($\times 10^{50}$~s$^{-1}$)} &  
 \colhead{($\times 10^{50}$~s$^{-1}$)} &
 \colhead{$\frac{Q_{\rm cluster}}{Q_{\rm H II}}^{\tablenotemark{a}}$}}
\startdata

1 & NGC\,5461-1  &       1\farcs3 & $0.47\pm0.05$ & $0.39\pm0.04$ & 
  $2\farcs1 \times 2\farcs0$ & 14 & 8 &  0.57 \\
2 & NGC\,5461-8,9,10   & 2\farcs2 & $0.43\pm0.08$ & $0.38\pm0.08$ & 
  $3\farcs5 \times 2\farcs8$ & 97 & 24\, & 0.25  \\
3 & NGC\,5462-18 &       1\farcs4 & $0.41\pm0.04$ & $0.35\pm0.04$ &
  $2\farcs4 \times 2\farcs3$ & 12 & 4 &  0.33  \\
4 & NGC\,5471-16 &       1\farcs5 & $0.30\pm0.03$ & $0.24\pm0.03$ &
  $2\farcs5 \times 2\farcs3$ & 37 & 8 &  0.22  \\

\enddata
\tablenotetext{a}{The ionizing luminosities of these clusters are 
 estimated assuming a cluster age of 3 Myr.  Since these clusters 
 are only know to have ages $<$ 5 Myr, the ionizing luminosity
 of the clusters and thus the ratio 
 $\frac{Q_{\rm cluster}}{Q_{\rm H II}}$ could range from 1/4 to
 1.5 times the value in the table.}
\end{deluxetable}

\end{document}